\documentclass[12pt,tightenlines,nofootinbib,amsmath,amssymb,aps,prd,notitlepage]{revtex4-1}

\usepackage{bm}
\usepackage{amssymb}
\usepackage[dvipsnames]{xcolor}
\usepackage{amsmath}
\usepackage{amsfonts}
\usepackage{mathtools}
\usepackage[english]{babel}
\usepackage{accents}
\usepackage{natbib}
\usepackage{enumitem}
\usepackage{natbib}
\usepackage{cancel}

\usepackage{hyperref}

\newcommand{\de}{\mbox{d}}
\newcommand{\be}{\begin{equation}}
\newcommand{\ee}{\end{equation}}
\newcommand{\pha}{\phantom{a}}

\newcommand{\pa}{\partial}

\newcommand{\sgn}{\mbox{sgn}}

\begin{document}

\title{A generalized Kasner transition for bouncing Bianchi~I models \\ \smallskip in modified gravity theories}

\author{Marco de Cesare}
\email{marco.de\_cesare@unb.ca}

\author{Edward Wilson-Ewing}
\email{edward.wilson-ewing@unb.ca}

\affiliation{Department of Mathematics and Statistics, University of New Brunswick, Fredericton, NB, Canada E3B 5A3}

\begin{abstract}

We derive transition rules for Kasner exponents in bouncing Bianchi~I models with generic perfect fluid matter fields for a broad class of modified gravity theories where cosmological singularities are resolved and replaced by a non-singular bounce.  This is a generalization of results obtained previously in limiting curvature mimetic gravity and loop quantum cosmology. A geometric interpretation is provided for the transition rule as a  linear map in the Kasner plane. We show that the general evolution of anisotropies in a Bianchi~I universe---including during the bounce phase---is equivalent to the motion of a point particle on a sphere, where the sphere is the one-point compactification of the Kasner plane. In addition, we study the evolution of anisotropies in a large family of bouncing Bianchi~I space-times.  We also present a novel explicit solution to the Einstein equations for a Bianchi~I universe with ekpyrotic matter with a constant equation of state $\omega=3$.

\end{abstract}

\maketitle

\section{Introduction}

In general relativity, if the matter content of the space-time satisfies the null energy condition, cosmological space-times include geodesics that are past-incomplete and therefore are singular \cite{Hawking:1969sw}; explicit example solutions show that space-time curvature invariants diverge at the singularity.  This singularity is typically understood to indicate that general relativity breaks down and must be replaced by a more fundamental theory that holds at higher energy/curvature scales than general relativity, and that is non-singular.

Among other possible resolutions, there exists a large variety of cosmological models where the initial singularity is resolved and is replaced by a non-singular bounce, specifically a regular space-time region connecting a contracting universe to an expanding universe.  A non-singular bounce can be generated in different ways, for example by quantum gravitational effects~\cite{Ashtekar:2011ni}, by introducing matter fields that violate the null energy condition, or due to modifications and/or extensions of the Einstein-Hilbert action~\cite{Brandenberger:2016vhg}.  Non-singular bouncing cosmological models are compatible with inflation, but have also been advocated as an alternative to the inflationary scenario \cite{Brandenberger:2016vhg}.

There is striking observational evidence for the homogeneity and isotropy of the universe on large scales, coming from high-precision measurements of the cosmic microwave background radiation (CMB), indicating that the cosmological background can be accurately described by the spatially homogeneous and isotropic Friedmann-Lema{\^i}tre-Roberston-Walker (FLRW) space-time.  However, (shear) anisotropies are expected to play an important role during the earliest stages of cosmic expansion, well before the CMB is produced, since their contribution to the equations of motion grows faster than most matter fields in a contracting space-time (or, equivalently, when evolving back in time in an expanding space-time).  Therefore, in models of the very early universe it is important to allow for the possibility of departures from the FLRW geometry, and in particular to include anisotropies.

Another reason to study anisotropic space-times comes from the Belinski-Khalatnikov-Lifshitz (BKL) conjecture in general relativity that claims that sufficiently near space-like singularities the Einstein equations become to a good approximation ultra-local \cite{Belinski:1982pk}, in which case the geometry can be effectively described by a metric of the Bianchi family at each point.  In particular, the Bianchi~VIII and Bianchi~IX space-times have the richest dynamics of the anisotropic (but homogeneous) space-times and therefore are the prototypes for constructing the general BKL solution.  Interestingly, these space-times exhibit a chaotic oscillatory behaviour (called Mixmaster) in the approach to the singularity.  The Mixmaster dynamics are characterised by an infinite sequence of so-called Kasner eras, during which the geometry is well approximated by a Bianchi~I space-time, marked by rapid transitions from one Bianchi~I solution to another, with these transitions generating the characteristic chaotic behaviour.  In the full inhomogenous BKL solution, even small curvature perturbations corresponding to differences in the initial conditions of neighbouring regions can lead to large inhomogeneities near the singularity, due to the fact that each region evolves chaotically and independently~\cite{Erickson:2003zm}: this phenomenon is known as the BKL instability.  This poses an important problem for bouncing cosmologies, since the BKL instability can potentially significantly increase inhomogeneities and destroy the smoothness of the bounce \cite{Brandenberger:2016vhg}.  (One way to avoid this problem is by introducing an ekpyrotic field with an equation of state $\omega>1$, which suppresses the anisotropies and ensures that the space-time remains isotropic at all times \cite{Erickson:2003zm}.)

For these reasons, within the context of cosmological scenarios based on a non-singular bounce, it is important to study the role of anisotropies, and in particular to understand how they propagate through the bounce. As a first step in this direction, we will consider the simplest anistropic space-time, namely the Bianchi~I space-time, in the framework of classical modified gravity theories.  For a large class of modified gravity theories, defined below, there is a simple map relating the pre-bounce Bianchi~I solution to the post-bounce Bianchi~I solution which can be expressed in terms of the Kasner exponents of the pre-bounce and post-bounce solutions; we call this map a generalized Kasner transition (since it is analogous but not identical to the Kasner transitions that occur in Bianchi~IX space-time between two consecutive Kasner eras).  This generalizes previous findings in limiting curvature mimetic gravity \cite{Chamseddine:2016uef} and loop quantum cosmology (LQC) \cite{Wilson-Ewing:2017vju}, and shows the much broader extent of their applicability.  Further, the generic occurrence of Kasner eras in all Bianchi space-times indicates that it may be possible to extend these results to the other Bianchi space-times as well.

In this paper, we will consider a class of modified gravity theories that resolve the cosmological singularity and replace it with a non-singular bounce.  (Note that a non-singular bounce can occur in cosmological space-times also in pure general relativity in the presence of positive spatial curvature and cosmological constant, although only for a somewhat restricted family of solutions \cite{Bramberger:2019oss}.  However, this is not possible for the spatially flat Bianchi I space-time.)

Specifically, we consider modified gravity theories that satisfy the following two conditions:
\begin{enumerate}[label=(\roman*)]
\item The departures from general relativity can be written in such a way that they are encoded in an effective stress-energy tensor,
\be \label{FieldEq}
G_{\mu\nu}= 8\pi G (T_{\mu\nu}+\tilde{T}_{\mu\nu})~,
\ee
where the gravitational coupling $G$ is constant, $T_{\mu\nu}$ is the usual stress-energy tensor, and $\tilde T_{\mu\nu}$ is the effective stress-energy tensor that captures the modifications to the Einstein equations.  Further, we assume that in the Bianchi~I space-time the matter field can be described as a perfect fluid and that (using the standard Cartesian spatial coordinates) the spatial components of $\tilde{T}^\mu{}_\nu$ are proportional to the identity, $\tilde{T}^i{}_j = -\tilde{p}\,\delta^i_j$.
\item In a cosmological setting (and in particular for the Bianchi~I space-time), while $T_{\mu\nu}$ may satisfy all energy conditions, the effective stress-energy tensor $\tilde T_{\mu\nu}$ violates the null energy condition sufficiently strongly and in such a way that the big-bang/big-crunch singularities of general relativity are avoided and are replaced by a non-singular bounce.  We also require that $\tilde{T}_{\mu\nu} \to 0$ away from the bounce so that the solutions of general relativity hold both far before and far after the bounce.
\end{enumerate}

Note that the last requirement in Condition (i) only imposes that the spatial components on the diagonal of $\tilde{T}^\mu{}_\nu$ must be identical---the symmetries of the Bianchi~I space-time alone already imply that $\tilde{T}^\mu{}_\nu$ is diagonal.  Another way to phrase this requirement is that $\tilde T_{\mu\nu}$ must have a perfect fluid form.  This implies that there is no preferred spatial direction in the effective stress energy tensor (not only at the level of the equations of motion, but also at the level of solutions to the equations for motion; for example, including a term proportional to the extrinsic curvature $K^i{}_j$ would not induce a preferred direction in $\tilde T_{\mu\nu}$ at the level of the equations of motion, but it would generate different contributions for the different entries of $\tilde T^\mu{}_\nu$ for any given specific anisotropic solution).

Also, the requirement in Condition (ii) that $\tilde T_{\mu\nu} \to 0$ away from the bounce can easily be relaxed (for example to include modified gravity theories where the effective $\tilde{T}_{\mu\nu}$ can also play the role of phantom dark energy at late times), we impose it here only for the sake of simplicity.

Many modified gravity theories satisfy Condition (i), some representative examples are:
\begin{itemize}
\item
In scalar-tensor theories, $\tilde{T}_{\mu\nu}$ represents the stress-energy of the scalar $\phi$. In a homogeneous cosmological space-time such as Bianchi~I, $\phi$ is a function of time only. Any minimally coupled scalar field, including for example a ghost field \cite{Barrow:2017yqt}, clearly satisfies Condition (i).  Further, isotropic contributions to $\tilde{T}_{\mu\nu}$ of the type $F g_{\mu\nu}$ are clearly compatible with Condition (i); here the function $F$ can include terms whose indices have been fully saturated to form scalar invariants, for example: the covariant Laplacian (or some powers thereof),  terms such as $\nabla_\sigma\phi\nabla^\sigma\phi$, $\nabla_\rho\nabla_\sigma\phi\nabla^\rho\nabla^\sigma\phi$, \dots , curvature invariants and contractions of the Riemann tensor with differential expressions involving $\phi$. On the other hand, in order to satisfy Condition (i), $\tilde T_{\mu\nu}$ cannot include most second-order covariant derivatives of the scalar field (or higher) with free indices; for example a term such as $\nabla^{\mu}\nabla_{\nu}\phi$ in $\tilde{T}^\mu{}_\nu$ would contribute terms proportional to the extrinsic curvature $K^i{}_j$ (which is, in general, not proportional to the identity) to $\tilde{T}^i{}_j$ and so are not included in the class of theories considered here. For similar reasons, $\tilde{T}_{\mu\nu}$ must not contain terms involving the Riemann tensor or the Ricci tensor with two unsaturated indices; thus, terms proportional to, e.g., $R_{\mu\nu}$, $R_{\mu\pha\nu}^{\pha\rho\pha\sigma} \nabla_{\rho} \phi \nabla_{\sigma} \phi$, \dots, which arise for instance in some Galileon theories \cite{Deffayet:2009wt}, are not allowed.  However, for terms such as $R^{\sigma}_{\pha(\mu} \nabla_{\nu)} \phi \nabla_{\sigma}\phi$, $R_{(\mu}^{\phantom{(}\pha\lambda\rho\sigma} \nabla_{\nu)} \phi \nabla_{\rho} \phi \nabla_{\lambda} \nabla_{\sigma}\phi$, where only one of the two unsaturated indices comes from the Riemann tensor, the only non-vanishing component is obtained for $\mu=\nu=0$; such terms are thus allowed, since they do not spoil the perfect fluid form of the effective stress-energy tensor.  In general, terms that in a cosmological context only contribute to $\tilde T_{00}$ are allowed.

These considerations can be directly generalized to the case of multiple scalar fields that depend only on time.

\item Other examples of scalar-tensor theories where Condition (i) is satisfied are the sub-class of Galileon theories with a minimally coupled scalar known as kinetic gravity braiding~\cite{Deffayet:2010qz, Easson:2011zy}, which can generate a non-singular bounce \cite{Lin:2010pf, Cai:2012va, Cai:2013vm}, as well as mimetic gravity~\cite{Chamseddine:2016uef}, which can also generate a non-singular bounce.

\item An effective gravitational coupling depending on a scalar field $\phi$ is typical of theories with non-minimal couplings to the curvature scalar $\phi R$ in the action, which generate a dynamical coupling $G_{\rm eff}$. For instance, in the general Brans-Dicke theory with a potential such a term is at the same time responsible for an effective gravitational coupling $16 \pi G_{\rm eff}=\phi^{-1}$, and for the presence of second-order operators with unsaturated indices in $\tilde{T}_{\mu\nu}$~\cite{Sotiriou:2011dz}. Nevertheless, Condition (i) is fully satisfied if the Brans-Dicke theory is reformulated in the Einstein frame. The situation is entirely analogous for metric $f(R)$ gravity (and also for Palatini $f(R)$ gravity), since this is equivalent to Brans-Dicke theory with vanishing Brans-Dicke parameter $\omega_{\rm BD}$ (or $\omega_{\rm BD}=-3/2$ for Palatini $f(R)$ gravity) \cite{Sotiriou:2006hs, Sotiriou:2011dz}.  As a result, the results derived in this paper hold for any Brans-Dicke theory expressed in the Einstein frame.  (Naturally, the results can be translated to the Jordan frame; this reformulation is presented in Appendix~\ref{FofRgravity}.)

\item In theories with a vector field, the vector field must point in the time direction.  As for non-minimal couplings and differential operators acting on the vector field, the same arguments applied to scalar-tensor theories can be extended to this case.  For example, Condition (i) is satisfied in Einstein-aether theory~\cite{Jacobson:2008aj}, provided the aether is hypersurface-orthogonal and the action functional only involves covariant derivatives of the aether through its divergence or a Maxwell-like kinetic term.

\item In ghost-free bimetric gravity \cite{Hassan:2011zd}, the coupling $G$ is necessarily constant since the kinetic term for each of the two metrics $g_{\mu\nu}$, $f_{\mu\nu}$ has the standard Einstein-Hilbert form. The other part of Condition (i) is also satisfied when both metrics $g_{\mu\nu}$ and $f_{\mu\nu}$ are aligned and have a line element that respects the symmetries of the Bianchi~I space-time (this is a generalization to the anisotropic sector of the isotropic cosmological ansatz for bimetric gravity \cite{vonStrauss:2011mq}), in which case it is straightforward to verify that $\tilde T_{\mu\nu}$ has a perfect fluid form in a Bianchi~I space-time%
\footnote{In this case the two line elements are $g_{\mu\nu}\de x^\mu\de x^\nu=\de t^2 - \bar{a}^2(t) (\sum_{i=1}^3 e^{2\beta_i(t)}\de x_i^2)$ and $f_{\mu\nu}\de x^\mu\de x^\nu=X^2(t)\de t^2 - Y^2(t) (\sum_{i=1}^3 e^{2\beta_i(t)}\de x_i^2)$. Hence the matrix $S = \sqrt{g^{-1}f \,}$ is diagonal and its spatial components are $S^i_{\pha j} = (Y(t)/\bar{a}(t)) \, \delta^i_{\pha j}$, where $Y$ and $\bar{a}$ are the mean scale factors. For both of these metrics, it is possible to write down a generalization of the Einstein equations of the form \eqref{FieldEq}, and the effective stress-energy tensor corresponding to the ghost-free interactions has the desired perfect fluid form.},
assuming matter only couples to one metric to avoid a Boulware-Deser ghost \cite{Schmidt-May:2015vnx}.

These considerations also apply to the de Rham-Gabadadze-Tolley (dRGT) theory of massive gravity \cite{deRham:2010kj, deRham:2010ik}, which arises as the limit of ghost-free bimetric gravity where one of the two metrics is non-dynamical.

\item There exist solutions for classical commuting spinor fields minimally coupled to general relativity, with self-interactions, that satisy Condition (i) in the Bianchi~I space-time~\cite{Saha:1996by}, one example being four-fermion interactions, which describe at an effective level the coupling between spin and space-time torsion in Einstein-Cartan-Sciama-Kibble theory and generalisations thereof \cite{Magueijo:2012ug}. While the presence of a spinor field in a Bianchi~I space-time constrains the space-time to be axially symmetric (with the anisotropic direction determined by the spinor axial current) \cite{Vignolo:2011qt}, the (space-like) axial current does not spoil the perfect fluid form of the stress-energy tensor since its only contribution to $\tilde T_{\mu\nu}$ is in the potential term (assuming no explicit couplings between the axial vector and other fields, e.g., vectors or derivatives of scalars) \cite{ArmendarizPicon:2003qk}. 

\end{itemize}
Not all modified gravity theories satisfy Condition (i), two counterexamples are the model derived from the 1-loop superstring effective action \cite{Kawai:1998bn} and the loop quantum cosmology effective dynamics \cite{Chiou:2007sp, Ashtekar:2009vc}.

The results derived in this paper hold for all theories satisfying Condition (i) with a non-singular bounce and general relativity as a low-curvature limit, thereby also satisfying Condition (ii).


\medskip

The outline of the paper is as follows. In Section~\ref{Sec2} we derive the conservation law for the densitized shear in Bianchi~I for all modified gravity theories that satisfy Condition (i), which plays a central role in the calculations in the following sections. Section~\ref{Sec3} contains some of the main results of this work, namely the transition rules relating the pre-bounce and post-bounce values for the Kasner exponents in the vacuum case and for a universe filled with matter with a stiff equation of state $\omega=1$. In Section~\ref{GeneralTransition}, by introducing time-dependent quasi-Kasner exponents it is possible to generalize these results to bouncing Bianchi~I space-times with arbitrary matter fields, and these quasi-Kasner exponents also can be used to obtain a geometric interpretation of the generalized Kasner transition produced by the bounce.  In Section~\ref{GeneralTransition} we also give what is to the best of our knowledge a novel exact solution to the Einstein equations for a Bianchi~I cosmology with ekpyrotic matter with equation of state $\omega=3$.  Our results are reviewed in the Conclusions.

Additional results are reported in the appendices. Appendix~\ref{App.Seq} describes the evolution of anisotropies in a general cosmological model with multiple matter fields, assuming a constant equation of state for each matter component, while Appendix~\ref{AppendixBounce} contains a detailed study of the evolution of anisotropies during the bounce and Appendix~\ref{MultiBounce} considers models with multiple bounces as well as the emergent universe scenario. Lastly, in Appendix~\ref{FofRgravity} the conservation law for the densitized shear in Bianchi~I cosmologies, for $f(R)$ gravity in the Einstein frame, is reformulated for the Jordan frame (this result can also be extended to the more general Brans-Dicke theory with a potential).

\section{Bianchi~I space-times in modified gravity}
\label{Sec2}

As stated in Condition (i) in Eq.~\eqref{FieldEq}, we consider modified gravity theories that satisfy the field equations $G_{\mu\nu} = T_{\mu\nu} + \tilde{T}_{\mu\nu},$ where $T_{\mu\nu}$ is the stress-energy tensor of ordinary matter, and $\tilde{T}_{\mu\nu}$ represents an {\it effective} stress-energy tensor which is responsible for deviations from general relativity.  Here and in the following, we have chosen units so that the (constant) gravitational coupling satisfies $8 \pi G = 1$.  $\tilde{T}_{\mu\nu}$ may include contributions from extra degrees of freedom, as well as higher-order derivatives of the metric; however, in order to avoid possible ambiguities we require that it does not include terms proportional to $G_{\mu\nu}$.  Note that while a constant gravitational coupling is assumed here, the equations of motion for many theories featuring a dynamical gravitational coupling (e.g., due to non-minimal couplings to curvature as in Brans-Dicke theory and its generalizations) can be written in the form of Eq.~\eqref{FieldEq} with a constant gravitational coupling by transforming to the Einstein frame.

It is convenient to recast Eq.~\eqref{FieldEq} as
\be
R_{\mu\nu}=\check{T}_{\mu\nu}-\frac{1}{2}\check{T}g_{\mu\nu} ~,
\ee
where the total stress-energy tensor is $\check{T}_{\mu\nu}=T_{\mu\nu}+\tilde{T}_{\mu\nu}$ and its trace is $\check{T}=g^{\mu\nu}\check{T}_{\mu\nu}$.

In the Bianchi~I space-time, the line element is
\be \label{LineElement}
\de s^2 =\de t^2 - \sum_{i=1}^3 a_i^2 \, \de x_i^2 ~,
\ee
with anisotropies present due to the different expansion rates of the directional scale factors~$a_i$.  Here $t$ is cosmic time and $x^i$ are comoving spatial coordinates. It is useful to define the mean scale factor as $\bar{a}=(a_1 a_2 a_3)^{1/3}$, and the proper volume of a unit comoving cell as $V=\bar{a}^3$.  The spatial metric is $h_{ij}= {\rm diag}(a_1^2, a_2^2, a_3^2)$, and its determinant is $h=V^2$. In the foliation adopted, the time-flow vector field and the unit normal to the spatial leaves coincide, i.e., $\frac{\pa}{\pa t}=n^\mu \frac{\pa}{\pa x^{\mu}}$, with $g_{\mu\nu}n^\mu n^\nu=1$.

For the Bianchi~I space-time, the modified Einstein equations reduce to \cite{Landau:1975}
\begin{align}
R^0_{\pha 0}&= -(\dot{K} + K_{ij}K^{ij})=\check{T}^0_{\pha 0}-\frac{1}{2}\check{T} ~,\label{TimeTimeFieldEq}\\
R^i_{\pha j}&=-\frac{1}{\sqrt{h}}\frac{\de}{\de t}(\sqrt{h} K^i_{\pha j})=\check{T}^i_{\pha j}-\frac{1}{2}\check{T}\delta^i_{\pha j} ~,\label{SpaceSpaceFieldEq} \\
R^0_{\pha i}&=D_j K^j_{\pha i}-D_i K =\check{T}^0_{\pha i} ~,\label{TimeSpaceFieldEq}
\end{align}
where a dot denotes a derivative with respect to the time coordinate $t$, $D_i$ is the covariant spatial derivative, and $K_{\mu\nu}\equiv \frac{1}{2}\mathcal{L}_{n} h_{\mu\nu}$ is the extrinsic curvature of constant $t$ hypersurfaces (and $K = h^{ij} K_{ij}$ is its trace), which for the Bianchi I space-time in these coordinates reduces to $K_{ij} = \frac{1}{2} \dot{h}_{ij}$.  Clearly, in the Bianchi~I space-time the extrinsic curvature depends only on time and therefore $\partial_i K_{jk} = 0$. Moreover, since the metric \eqref{LineElement} is spatially flat, the spatial components of the Christoffel symbols are all zero. As a result, $D_i K_{jk} = 0$ and so $\check{T}^0_{\pha i}=0$.  As a result, the dynamics is governed entirely by Eqs.~\eqref{TimeTimeFieldEq}--\eqref{SpaceSpaceFieldEq}. Following Ref.~\cite{Chamseddine:2016uef}, subtracting from Eq.~\eqref{SpaceSpaceFieldEq} one third of its trace gives
\be \label{SpaceSpace_TraceFree}
-\frac{1}{\sqrt{h}}\frac{\de}{\de t}\left[\sqrt{h}\left( K^i_{\pha j}-\frac{1}{3}K \delta^i_{\pha j}\right)\right]=\check{T}^i_{\pha j}-\frac{1}{3}\check{T}^k_{\pha k}\delta^i_{\pha j} ~.
\ee
This can be further simplified by splitting the extrinsic curvature into its trace and shear, $K_{ij}=\frac{1}{3}K h_{ij}+\sigma_{ij}$ where $\sigma_{ij}$ is the (trace-free) shear tensor.  In the coordinate system~\eqref{LineElement} where $h_{ij}$ is diagonal, $K_{ij}$ is necessarily also diagonal, and therefore the shear tensor is diagonal as well, i.e., $\sigma_{ij}=\sigma_{(i)}h_{ij}$ (no sum over $i$), where $\sigma_{(i)}$ denote the three entries on the diagonal in this coordinate system. Then, Eq.~\eqref{SpaceSpace_TraceFree} implies that $\check{T}^i{}_j$ must also be diagonal in this coordinate system.  Denoting the principal pressures by $\check{p}_i$, $\check{T}^i_{\pha j}=-\check{p}_i\delta^i_{\pha j}$~. Of particular interest is the case in which all of the principal pressures coincide, i.e., $\check{p}_i=\check{p}$.  The contribution of any homogeneous perfect fluid matter fields (including a scalar field) to $T_{\mu\nu}$ will satisfy this condition, and Condition (i) requires that $\tilde T^i{}_j = - \tilde p \, \delta^i_j$.  Therefore, for the class of theories we are interested in here, the dynamical equations~\eqref{TimeTimeFieldEq}--\eqref{SpaceSpaceFieldEq} do not single out a preferred direction and Eq.~\eqref{SpaceSpace_TraceFree} simplifies to a conservation law for the densitized shear,
\be \label{ConservationLaw}
\frac{\de}{\de t} \left(\sqrt{h}\, \sigma^i_{\pha j}\right)=0~.
\ee
This shows that the conservation law~\eqref{ConservationLaw}, previously obtained in general relativity \cite{Montani:2007vu} and in mimetic gravity~\cite{Chamseddine:2016uef}, holds for a much larger class of theories.

Anisotropies in the universe evolve according to Eq.~\eqref{ConservationLaw}, whose solution gives the first integral
\be\label{SolShear}
\sigma^i_{\pha j}=\frac{\Lambda^i_{\pha j}}{\sqrt{h}} ~,
\ee
where $\Lambda^i_{\pha j}$ is a constant symmetric matrix with vanishing trace, $\Lambda^i_{\pha i}=0$; its entries are determined by the choice of initial conditions for $a_i$ and $\dot a_i$. The effective Friedmann and Raychaudhuri equations for the evolution of the mean scale factor $\bar{a}$ can be derived from the field equations \eqref{FieldEq}
\begin{align}
\bar{H}^2&=\frac{1}{3}\left(\check{\rho}+\frac{\Sigma^2}{h}\right) ~,\label{Friedmann_Mean}\\
\dot{\bar{H}}&=-\frac{1}{2}\left(\check{\rho}+\check{p}+\frac{2 \Sigma^2}{h}\right) ~,\label{Raychaudhuri_Mean}
\end{align}
where $\bar{H}=\dot{\bar{a}}/\bar{a}$ is the mean Hubble rate, $\check{\rho}=\check{T}^0{}_0 = T^0{}_0 + \tilde T^0{}_0 = \rho + \tilde \rho$ is the total energy density, and
\be \label{def-shear}
\Sigma^2\equiv \frac{1}{2}\Lambda^i{}_j\Lambda^j{}_i = \frac{h}{6}\Big[(H_1 - H_2)^2 + (H_1 - H_3)^2 + (H_2 - H_3)^2\Big]
\ee
is the shear scalar which is a constant of the motion, and $H_i = \dot a_i / a_i$ are the directional Hubble rates.  The shear contributes to the effective Friedmann equation \eqref{Friedmann_Mean} a term proportional to $\bar{a}^{-6}$; this is the same scaling behaviour characterising stiff matter ($\omega=1$).  It is often convenient to introduce
\be \label{defOmega}
\Omega_\Sigma = \frac{\Sigma^2}{3 \bar{H}^2 \bar{a}^6}~,
\ee
which measures the fractional contribution of the shear to the Friedmann equation~\eqref{Friedmann_Mean}.  Note that in general $\Omega_\Sigma$ is not a constant of the motion (except in vacuum or if the only matter content is a stiff fluid).

The dynamics is fully specified by Eqs.~\eqref{Friedmann_Mean}, \eqref{Raychaudhuri_Mean} and the first integral \eqref{SolShear}.  Note that, despite the familiar form in which these equations have been recast, these are {\it not} the standard Friedmann and Raychaudhuri equations of general relativity. In fact, the effective energy density and pressure include contributions from $\tilde T^\mu{}_\nu$ arising from modifications of the Einstein equations (or modifications to the Einstein-Hilbert action if the theory is formulated starting from an action principle), and typically lead to departures from general relativity which can be significant in certain regimes.

\section{Bouncing Bianchi~I space-time in modified gravity}
\label{Sec3}

Following Condition (ii) on the class of modified gravity theories considered here, we shall restrict our attention to those theories in which Friedmann and Kasner singularities are resolved.  For this to be possible, the total stress-energy tensor $\check{T}_{\mu\nu}$ must violate the null energy condition (NEC) in order to produce a bounce in a spatially flat universe filled with ordinary matter and anisotropies.  In the following, we shall assume that ordinary matter satisfies the energy conditions, and that the NEC violation is entirely due to the effective stress-energy tensor $\tilde{T}_{\mu\nu}$ coming from the modification of the Einstein equations. Thus, in a neighbourhood of the bounce
\be \label{NECviolation}
\check{\rho}+\check{p}=\tilde{\rho}+\tilde{p}+\rho+p<0~\implies~~\tilde{\rho}+\tilde{p}<0~,
\ee
with $\rho,\,\rho+p\geq0$. NEC violation \eqref{NECviolation} implies that the Hubble rate increases from negative values in the pre-bounce phase to positive values in the post-bounce one. The effective fluid must also violate positivity of the energy density {\it at} the bounce, whereby the mean Hubble rate vanishes and therefore, since $\Sigma^2 \ge 0$,
\be \label{WECviolation}
\check{\rho}=\tilde{\rho}+\rho\le 0~\implies~~\tilde{\rho}\le -\rho<0~.
\ee
Thus, by continuity the effective energy density $\tilde{\rho}$ must be negative in a neighbourhood of the bounce.  Finally, as already mentioned in Condition (ii) above, we also require that $\check{T}_{\mu\nu} \to T_{\mu\nu}$, or equivalently $\tilde \rho \to 0$ and $\tilde p \to 0$, away from the bounce so that the solutions of general relativity hold both far before and far after the bounce.

\subsection{Generalized Kasner transition in vacuo or with a stiff fluid}
\label{ZeldovichUniverse}

The simplest case is when the matter field is a massless scalar field (or more generally, a stiff fluid), in which case the equation of state is $\omega=p/\rho=1$, or the vacuum case when there is no matter.  Since vacuo is a limiting case of a stiff fluid, these two cases can be considered together.  Other types of perfect fluid matter content are considered in Section~\ref{bounce.gen.matter}.

The dynamics of a Bianchi~I space-time with a stiff fluid is particularly simple because the energy density and pressure of the stiff matter scales in the same way as the contribution of the anisotropies to Friedmann and Raychaudhuri equations.  Since stiff matter was first considered by Zel'dovich~\cite{Zeldovich:1962}, following Jacobs~\cite{Jacobs_1968} we will refer to a Bianchi~I space-time with stiff matter satisfying the Einstein equations of general relativity as a Zeld'ovich universe.

Under the above assumptions, the bounce can be regarded as a transition from a contracting Zeld'ovich universe to an expanding one, mediated by a bounce during which the modifications to the Einstein equations, captured in $\tilde{T}_{\mu\nu}$, become important.  As explained above, away from the bounce $\tilde{T}_{\mu\nu} \to 0$ and the dynamics are determined by the Einstein equations.  As shall be shown here, the Kasner exponents parameterizing the pre- and post-bounce Zeld'ovich universes are related in a simple way determined by the conservation law~\eqref{ConservationLaw} and the effective Friedmann equation~\eqref{Friedmann_Mean}.

For convenience, we set the origin of time $t=0$ at the bounce without any loss of generality.  In a Zeld'ovich universe, the Einstein equations imply that the volume $V$ evolves linearly with time
\be \label{VolumeSolStiff}
V_\pm(t)=V^{0}_{\pm}+\upsilon_{\pm} |t| ~.
\ee
Clearly, this solution will hold far from the bounce when the modifications to general relativity are negligible.  Here the subscript `$-$' corresponds to the contracting pre-bounce solution and `$+$' denotes the expanding post-bounce solution, while $V^0_\pm$ are two constants of integration and
\be
\upsilon_{\pm} = \sqrt{3\left(\Sigma^2+\frac{({\pi_\phi}^{\pm})^2}{2}\right)}~
\ee
is also constant in a Zeld'ovich universe.  The constant $\Sigma^2$ is defined in \eqref{def-shear}, while $\pi_\phi^{\pm}$ is the momentum of the massless scalar field, another constant of the motion in the regime away from the bounce where the Einstein equations hold.  (More generally, for a generic stiff fluid $\pi_\phi^2/2$ can be replaced by the constant of the motion $\rho V^2$.)  Due to \eqref{ConservationLaw}, $\Sigma^2$ is necessarily the same on either side of the bounce, and in fact remains constant throughout the bounce, but this is not necessarily the case for $\pi_\phi$, so it is important to allow for the possibility that $\pi_\phi$ may change during the bounce due to the modifications of the Einstein equations.  Of course, away from the bounce, in either of the Zeld'ovich universes, each $\pi_\phi^\pm$ is constant.  The vacuum limit is obtained for $\pi_\phi=0$.

It is convenient to introduce the variables $\alpha_i=\log a_i$, whose first time derivative gives the directional Hubble rates $H_i$.  From the decomposition of the extrinsic curvature in terms of expansion and shear,
\be \label{eom-alpha}
\dot{\alpha}_i=\frac{1}{3}\frac{\dot{V}}{V}+\frac{\lambda_i}{V} ~,
\ee
where $\lambda_i$ denote the eigenvalues of the tensor $\Lambda^i_{\pha j}$ (which is constant and diagonal in the coordinates used here) in Eq.~\eqref{SolShear}. Integrating this equation gives
\be \label{SolAlpha}
\alpha_i=\alpha_i^0+\frac{1}{3}\log V+\int\de t\;\frac{\lambda_i}{V} ~,
\ee
where the integration constants satisfy $\sum_{i=1}^3\alpha_i^0=0$.  Importantly, Eq.~\eqref{SolAlpha} is valid at all times including the bounce.  Substituting Eq.~\eqref{VolumeSolStiff} into Eq.~\eqref{SolAlpha} gives the solution for the pre- and post-bounce Zeld'ovich universes,
\be \label{SolAlphawKasner}
\alpha_i = \alpha_{i\,\pm}^0 + k_i^{\pm} \log V ~,
\ee
where the Kasner exponents are $k_i^{\pm}=\frac{1}{3}\pm\frac{\lambda_i}{\upsilon_{\pm}}$. The field equations imply that the Kasner exponents must satisfy the following relations \cite{Wilson-Ewing:2017vju, Erickson:2003zm}
\be \label{ZeldovichRels}
\sum_{i=1}^3 k_i^{\pm}=1 ~, \hspace{1em}
\sum_{i=1}^3 (k_i^{\pm})^2=1-(k_\phi^{\pm})^2 ~,
\ee
where $k_\phi^{\pm}=\pi_\phi^{\pm}/\upsilon^{\pm}$.

From the conservation law~\eqref{ConservationLaw} and the definition of $\sigma^i{}_j = h^{ik} \sigma_{kj}$ as the trace-free part of $K^i{}_j$,
\be \label{ConservationLawwAlpha}
V\bigg(\dot{\alpha}_i-\frac{1}{3}\sum_j \dot{\alpha}_j\bigg)= {\rm constant} ~.
\ee
Importantly, this relation holds at all times including the bounce, and does not depend on the matter content beyond the requirements listed in Condition (i).  Differentiating \eqref{SolAlphawKasner} gives $\dot{\alpha}_i = \pm\upsilon_{\pm} V^{-1} k_i^{\pm}$, and summing over $i$ results in $\sum_i \dot{\alpha}_i=\pm\upsilon_{\pm}V^{-1}$.  Combining these results, evaluated in the pre-bounce and post-bounce Zeld'ovich universes, together with \eqref{ConservationLawwAlpha} gives
\be
\upsilon_{+}\left(k_i^{+}-\frac{1}{3}\right)=-\upsilon_{-}\left(k_i^{-}-\frac{1}{3}\right) ~.
\ee
This relation can be rewritten to obtain the transition rule
\be \label{TransitionRule}
k_i^{+}=\frac{1}{3}\left(1+\frac{\upsilon_{-}}{\upsilon_{+}}\right)-\left(\frac{\upsilon_{-}}{\upsilon_{+}}\right)k_i^{-} ~,
\ee
which relates the Kasner exponents in the pre- and post-bounce Zeld'ovich universes; we call this transition rule a generalized Kasner transition. Note that this transition rule is linear in the Kasner plane, and $k_i^+$ depends only on $k_i^-$---the transformation of $k_i$ is independent of the values of the other two Kasner exponents.  These two properties of the generalized Kasner transition were not assumed but rather are the result of the calculation above.

If the momentum of the scalar field is conserved $\pi_\phi^{+}=\pi_\phi^{-}$, then $\upsilon_{+}=\upsilon_{-}$ and the same transition rule found for limiting curvature mimetic gravity is recovered \cite{Chamseddine:2016uef},
\be \label{TransitionRuleSymm}
k_i^{+}=\frac{2}{3}-k_i^{-} ~.
\ee

In some cases, a transition rule for the Kasner exponents can be derived even if the densitized shear $\Sigma^2$ is not constant during the bounce.  For example, in loop quantum cosmology $\Sigma^2$ is not constant during the bounce (although it does remain finite), but $\Sigma^2$ does asymptote to the same constant value either side of the bounce \cite{Chiou:2007sp, Ashtekar:2009vc}.  In this case, it can be shown that the transition rule \eqref{TransitionRuleSymm} also holds, although the proof follows a different path \cite{Wilson-Ewing:2017vju}.

\subsection{Geometric interpretation of the generalized Kasner transition}
\label{Sec_GeoStiff}

\begin{figure}[t]
\begin{center}
\includegraphics[width=0.5\columnwidth]{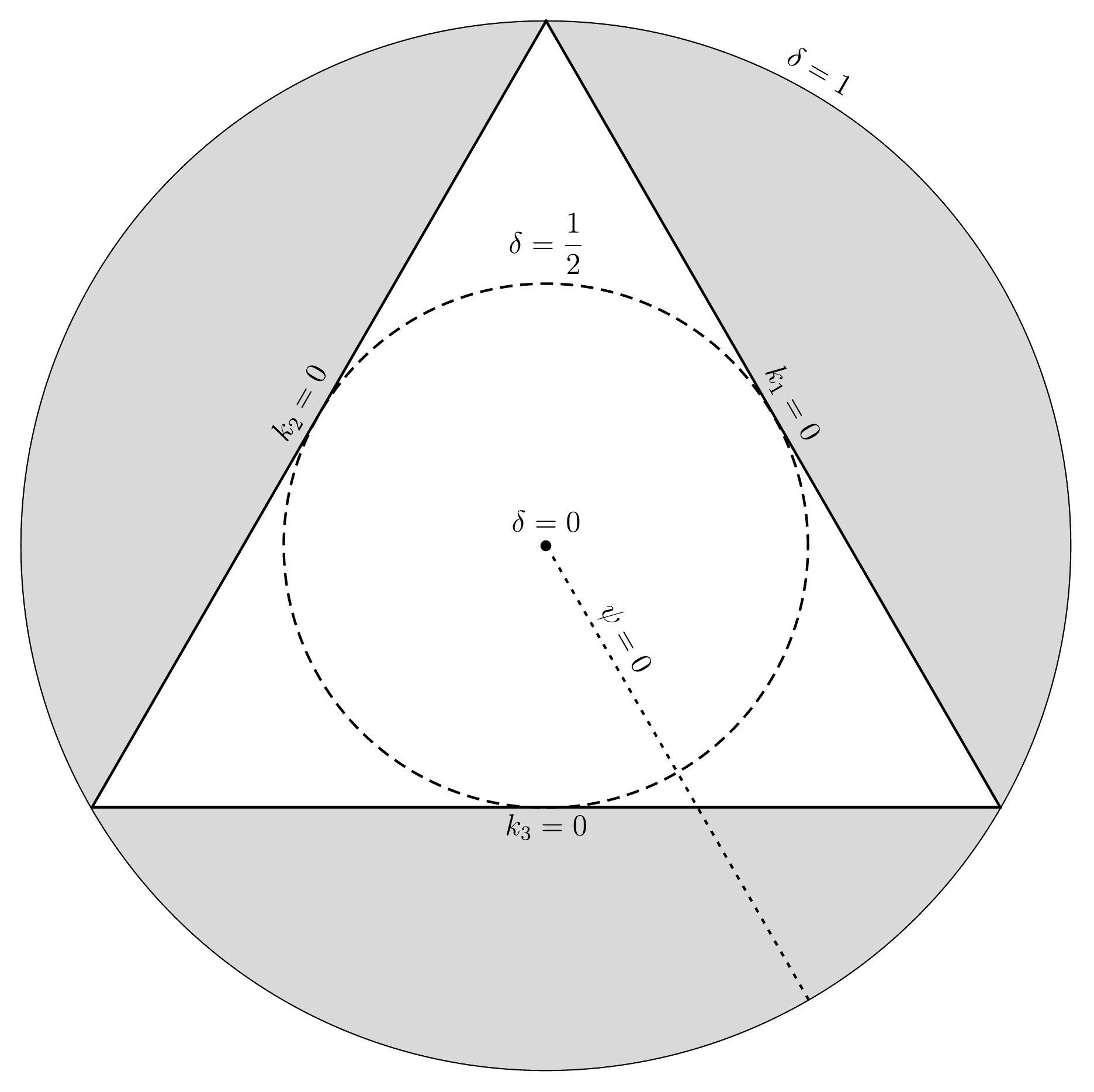}
\caption{\hangindent=12pt \small
This figure represents the Kasner disc (see also Ref.~\cite{Erickson:2003zm}). In general relativity, different regions of the Kasner disc correspond to  different types of solutions.  The Kasner circle is the boundary of the disc where $\delta=1$ and represents the set of vacuum Bianchi~I solutions in general relativity. Inside the gray region there is exactly one negative Kasner exponent, whereas inside the triangle all Kasner exponents are positive. The sides of the triangle are characterised by one vanishing Kasner exponents, and the vertices correspond to space-times that are diffeomorphic to Minkowski. The center of the disc corresponds to a flat FLRW universe. The Jacobs parametrisation $(\delta,\psi)$ of the disc is also illustrated. Note that the radial parameter $\delta$ is a measure of the anisotropies, with $\delta=0$ corresponding to perfect isotropy and $\delta=1$ to maximal anisotropy, which arises in a vacuum Kasner space-time.} \label{KasnerDisc} 
\end{center}
\end{figure}

The Zeld'ovich universe is fully characterised by the parameters $k_i$, $k_\phi$, related by Eq.~\eqref{ZeldovichRels} (dropping for now the indices denoting contracting and expanding branches). These algebraic relations can be plotted in a three-dimensional Euclidean space with the Cartesian coordinates $k_i$, where they represent the circle obtained as the intersection of the Kasner plane $\sum_{i=1}^3 k_i=1$ with the sphere centered in the origin with radius $r^2 = 1-k_\phi^2$ \cite{Erickson:2003zm}; this is shown in Fig.~\ref{KasnerDisc}. The vacuum case is obtained for $k_\phi=0$, and the corresponding circle of radius $\sqrt{2/3}$ is referred to as the Kasner circle. In the opposite limit case, namely $k_\phi=1$, the circle shrinks to a point and the universe is isotropic (i.e., spatially flat FLRW). Thus, all solutions of the Einstein equations for Bianchi I dominated by stiff matter can be mapped into the planar disc bounded by the Kasner circle.  (There are three special cases, namely the vertices of the triangle in Fig.~\ref{KasnerDisc}, which are diffeomorphic to Minkowski space-time.)

For a planar domain, it is convenient to introduce an alternative parametrisation based on two independent real parameters.  One choice is \cite{Jacobs_1968,Gupt:2012vi}
\be\label{JacobsParametrization}
k_i=\frac{1}{3}\Big(1+2\delta\, Z_i(\psi)\Big) ~,\hspace{1em}
k_\phi^2= \frac{2}{3}(1-\delta^2) ~.
\ee
where $0\leq\psi<2\pi$, $0\leq\delta\leq1$ and $Z_n(\psi)=\sin[\psi+\frac{2\pi}{3}(n-1)]$ (note that the range of the variables defined here is different to the choice made in \cite{Jacobs_1968, Gupt:2012vi}).  Constant $\psi$ corresponds to a half-line in the Kasner plane with the endpoint $k_1 = k_2 = k_3 = \frac{1}{3}$, and constant $\delta$ is a circle in the Kasner plane with radius $r^2 = 2\delta^2/3$ (not to be confused with the radius $R^2=\frac{1}{3} + \frac{2}{3} \delta^2$ of the sphere satisfying the $\sum_{i=1}^3 k_i^2 = 1 - k_\phi^2$ constraint in the three-dimensional space $(k_1, k_2, k_3)$; it is the intersection of this sphere with the Kasner plane $\sum_{i=1}^3k_i = 1$ that gives a circle with a radius $r^2 = 2\delta^2/3$).

The radius of the circle is completely determined by the contribution of anisotropies to the energy density,
\be \label{delta-stiff}
\delta^2=\frac{\Sigma^2}{\,\frac{1}{2}\pi_\phi^2+\Sigma^2}=\Omega_{\Sigma}~,
\ee
with $\Omega_\Sigma$ defined in~\eqref{defOmega}.

The transition rule \eqref{TransitionRule} for the bounce between two asymptotic Zeld'ovich space-times implies
\be \label{TransitionRuleAngles}
Z_1(\psi^{+}) = Z_1(\psi^{-}+\pi) ~,~ Z_2(\psi^{+}) = Z_2(\psi^{-}+\pi)~,~ Z_3(\psi^{+}) = Z_3(\psi^{-}+\pi) ~,
\ee
geometrically this sends the Kasner exponents from the half-line $\psi^-$ to the half-line $\psi^+ = \psi^- + \pi$.  If the scalar field momentum is conserved across the bounce, the Kasner exponents for both expanding and contracting phases lie on the same circle with $\delta_{+}=\delta_{-}$, and the transition rule, expressed in different equivalent forms by Eqs.~\eqref{TransitionRuleSymm} and~\eqref{TransitionRuleAngles}, is the antipodal map on the circle. In the general case, the transition rule \eqref{TransitionRule} corresponds to a composition of the antipodal map~\eqref{TransitionRuleAngles} and the rescaling of the radius of the circle $\delta_{-}\to\delta{+}$, with
\be
\delta_\pm^2=\frac{\Sigma^2}{\,\frac{1}{2}(\pi_\phi^\pm)^2+\Sigma^2}.
\ee

\section{Quasi-Kasner exponents}
\label{GeneralTransition}

\subsection{Definition of quasi-Kasner exponents}

To study Bianchi~I space-times with other types of perfect fluid matter (i.e., with an equation of state $\omega = p / \rho \neq 1$), we introduce time-dependent quasi-Kasner exponents defined as\footnote{Note that this definition holds very generally.  For modified gravity theories that satisfy Condition (i) $\lambda_i$ will necessarily be a constant of the motion, but in other modified gravity theories $\lambda_i$ may be dynamical.  In both cases this definition of quasi-Kasner exponents holds.}
\be \label{QuasiKasner}
k_i(t) = \frac{1}{3 \bar H} \cdot \frac{\dot a_i}{a_i} = \frac{1}{3}+\frac{\lambda_i}{\dot{V}} ~.
\ee
From \eqref{eom-alpha} and the definition of the time-dependent quasi-Kasner exponents, it follows that
\be
\dot{\alpha}_i=k_i\frac{\dot{V}}{V}~,
\ee
and the directional scale factors are therefore given by
\be
a_i=a_i^0\exp\left(\int_{t_0}^t \de t\; k_i\bar{H} \right) ~.
\ee
In a Zeld'ovich universe the volume evolves linearly with time \eqref{VolumeSolStiff} and the standard definition of the Kasner exponents is recovered.

The time-dependent Kasner exponents obey the following algebraic relations
\be \label{quasi-kasner-const}
\sum_{i=1}^3 k_i=1 ~, \hspace{1em}
\sum_{i=1}^3 k_i^2=\frac{1}{3}\left(1+2\,\Omega_{\Sigma}\right) ~,
\ee
generalizing the usual relations that hold in a Zeld'ovich universe.

The Jacobs parametrisation, given by the first equation in \eqref{JacobsParametrization}, still holds although with a time dependent $\delta = \sqrt{\Omega_{\Sigma}}$. From the definition~\eqref{QuasiKasner}, the time-dependent quasi-Kasner exponents satisfy the differential equation
\be \label{EqForK_general}
\dot{k}_i+3\bar{H} \left(k_i-\frac{1}{3}\right)=\frac{\epsilon}{3}\frac{\lambda_i}{V} ~,
\ee
where $\epsilon=-\dot{\bar{H}}/\bar{H}^2$ (in the context of inflationary models, $\epsilon$ is a slow-roll parameter).  The $\epsilon$ parameter will be used in this section for convenience, in order to describe a universe filled with generic matter, including the case of a time-varying equation of state. If the dynamics of the Bianchi~I space-time is dominated by a matter field with an equation of state parameter $\omega = p/\rho$, then $\epsilon=\frac{3}{2}(\omega+1)$.

From Eq.~\eqref{EqForK_general}, $\sum_{i=1}^3 \dot{k}_i=0$, which shows that the quasi-Kasner exponents are constrained to the Kasner plane. Moreover, from \eqref{quasi-kasner-const} and the definition of $\Omega_{\Sigma}$ in \eqref{defOmega}, it follows that
\be \label{ODE_CircleRadKasnerPlane1}
\frac{\de}{\de t}\left(\sum_{i=1}^3 k_i^2\right)=-4\left(1-\frac{\epsilon}{3}\right)\bar{H} \Omega_{\Sigma} ~,
\ee
which can be rewritten as
\be \label{ODE_CircleRadKasnerPlane2}
\frac{\de}{\de t}\log\delta=-(3-\epsilon)\bar{H} ~.
\ee
This relation shows that, unlike the Kasner and Zeld'ovich solutions (for which $\epsilon=3$), in the general case the Kasner exponents do not single out a circle with fixed radius in the Kasner plane. In fact, the evolution of the radial parameter depends on the equation of state of the dominant matter component, as well as on $\bar{H}$ (and particularly on its sign). If the universe is contracting, anisotropies grow in time for matter with $\epsilon<3$, while they decrease for ekpyrotic matter with $\epsilon>3$; the situation is reversed if the universe is expanding.

In terms of the quasi-Kasner exponents, in an expanding Bianchi~I space-time in general relativity with ordinary matter (i.e., with equation of state $\omega<1$ and $\epsilon<3$), the space-time will become isotropic with $\Omega_\Sigma \to 0$ and all quasi-Kasner exponents tending to $1/3$, while near the singularity $\Omega_\Sigma \to 1$ and the Kasner exponents will lie on the Kasner circle.  On the other hand, in an expanding Bianchi~I space-time with ultra-stiff, or ekpyrotic, matter (i.e., with equation of state $\omega>1$ and $\epsilon > 3$), it is near the singularity that the space-time is nearly isotropic, with all quasi-Kasner exponents $1/3$ and $\Omega_\Sigma \to 0$, while at late times $\Omega_\Sigma \to 1$ and the Kasner exponents lie on the Kasner circle.

The case $\epsilon=3$ corresponding to stiff matter, where $\dot\delta=0$, has been analyzed in Section~\ref{ZeldovichUniverse}.

Further, by using $\bar H = \dot V/3V$ and the definition of $\Omega_\Sigma$, the equation~\eqref{QuasiKasner} can be re-expressed as
\be \label{KasnerEvolve}
k_i(t)=\frac{1}{3}\left(1+\lambda_i\,\sgn( \bar{H}) \sqrt{\frac{3 \Omega_{\Sigma}}{\Sigma^2}\,} \right) ~,
\ee
and in terms of the Jacobs parameterization~\eqref{JacobsParametrization},
\be \label{AngularVarTransition}
Z_i(\psi) = \sqrt \frac{3}{4\Sigma^2} \, \lambda_i\,\sgn( \bar{H}) ~.
\ee
Unlike $\delta$, $Z_i$ is constant (and therefore the Jacobs angle $\psi$ is also constant) both in the contracting phase and the expanding one, although it is discontinuous through the bounce (in fact, as shall be shown below, the bounce relates the Jacobs angle before and after the bounce by $\psi^+ = \psi^- + \pi$). Assuming the dynamics is dominated by matter fields with equations of state $\omega < 1$, then the quasi-Kasner exponents of a contracting Bianchi~I space-time stay on the half-line $r^-$ given by constant $\psi^-$ and move in the direction of increasing $\delta$; similarly, for the expanding Bianchi~I space-time the quasi-Kasner exponents stay on the half-line $r^+$ of constant $\psi^+$, moving in the direction of decreasing $\delta$, and will asymptote to the isotropic point $k_1 = k_2 = k_3 = \frac{1}{3}$ that lies at the end of the half-line. (For an ekpyrotic matter field, the behaviour is reversed between expanding and contracting Bianchi~I space-times, while for a stiff fluid the Kasner exponents are constant.)

Independently of the matter content of the Bianchi~I space-time, if $\sgn(\bar H)$ does not change then the quasi-Kasner exponents always remain on the same half-line of constant $\psi$.

\subsection{Examples}
\label{exact-sols}

To gain a better understanding of the quasi-Kasner exponents, it is helpful to consider two exact solutions in general relativity for a Bianchi~I space-time filled with hydrodynamic matter, one with a stiff fluid and dust (i.e., $p=0$), and one with a stiff fluid and an ekpyrotic perfect fluid with a constant equation of state $\omega=3$.  Beyond these two examples with exact explicit solutions, the case of a Bianchi~I space-time with multiple matter fields is considered in Appendix~\ref{SharpTransitions}.

\subsubsection{Heckmann-Sch\"ucking solution}

The Heckmann-Sch\"ucking solution describes a Bianchi~I universe filled with a stiff fluid ($\omega=1$) and dust ($\omega=0$), the directional scale factors evolve as \cite{Heckmann:1962, Khalatnikov:2003ph, Ali:2017qwa}
\be\label{HSsolution}
a_i(t)=a_0\,t^{k_{i,0}}(t+t_0)^{\frac{2}{3}-k_{i,0}} ~.
\ee
The singularity occurs at $t=0$, where the energy density is dominated by the stiff fluid and anisotropies. In this limit, the Kasner exponents are $k_{i,0}$ and necessarily lie within the Kasner disc.  The time-dependent quasi-Kasner exponents are given by
\be
k_i(t)=\frac{ k_{i,0}+\frac{2t}{3t_0}}{1+\frac{2t}{t_0}} ~,
\ee
and the density parameter for the anisotropies is
\be
\Omega_{\Sigma}(t)=\left(\frac{t_0}{2t+t_0}\right)^2 \Omega_{\Sigma,0}~,
\ee
with the initial condition for anisotropies satisfying $\sum_i k_{i,0}^2=\tfrac{1}{3}(1+2\Omega_{\Sigma,0})$.  Recalling that the Jacobs angle $\psi$ is constant and that the Jacobs parameter $\delta$ is given by $\delta = \sqrt{\Omega_\Sigma}$, as $t$ increases the point in the Kasner plane moves from its starting point (at the initial singularity) somewhere within the Kasner disc to the central isotropic point in a straight line.

\subsubsection{Universe with a stiff fluid and ekpyrotic matter}
\label{ekpyrotic-sol}

Another exact solution can be found for a Bianchi~I space-time with a stiff fluid and ekpyrotic matter with $\omega=3$,
\be\label{EkpyroticExactSol}
a_i(t)=a_0\,t^{\frac{1}{6}}(t+t_0)^{\frac{1}{6}}\left(\sqrt{t}+\sqrt{t+t_0}\right)^{2k_{i,0}-\frac{2}{3}} ~,
\ee
to the best of our knowledge, this exact solution to the Einstein equations has not been explicitly written down before (although it is implicitly included in the family of solutions considered in Ref.~\cite{Coley1992} where the asymptotic early- and late-time dynamics of Bianchi space-times with two perfect fluids are studied).  Again, the singularity occurs at $t=0$ and the space-time expands as $t$ increases.  Now, the $k_{i,0}$ are the Kasner exponents at late times (and again necessarily lie within the Kasner disc).

In this case, the quasi-Kasner exponents are given by
\be
k_i(t)=\frac{1}{3}+\frac{2\sqrt{t(t+t_0)}}{2t+t_0} \left(k_{i,0}-\frac{1}{3}\right) ~.
\ee
Compared to the Heckmann-Sch\"ucking solution, the evolution in the Kasner plane follows an opposite trend: starting from an anisotropic universe for large $t$, an isotropic configuration is reached as $t$ tends to zero. The density parameter for anisotropies is given by
\be
\Omega_{\Sigma}=\frac{4t(t+t_0)}{(2t+t_0)^2}\Omega_{\Sigma,0}~,
\ee
from which the evolution of $\delta$ can be obtained. Here $\Omega_{\Sigma,0}$ represents the density parameter for anisotropies in the limit $t\to+\infty$.  Again, the Jacobs angle $\psi$ is constant, so the quasi-Kasner exponents evolve in a straight line in the Kasner plane, starting from the isotropic point $k_1 = k_2 = k_3 = \frac{1}{3}$ and moving to a limit point within the Kasner disc with radial Jacobs parameter $\delta_0=\sqrt{\Omega_{\Sigma,0}}$.

\subsection{Application to modified gravity theories with a bounce}
\label{bounce.gen.matter}

In the modified gravity theories considered here, during the bounce the NEC must be violated by the total stress-energy tensor $\check T_{\mu\nu}$ during which time $\epsilon<0$. The effective fluid must also violate positivity of the energy density and therefore (unless the shear scalar $\Sigma$ is exactly zero) the inequality $\Omega_{\Sigma}\leq 1$ must also be violated during the bounce. In fact, at the bounce point $\bar{H}\to 0$ while the scale factor is finite and therefore $\Omega_{\Sigma}$ diverges. Since $\delta = \sqrt{\Omega_{\Sigma}}$, the quasi-Kasner exponents will necessarily exit the Kasner circle $\delta=1$ during the bounce, and in fact reach an infinite value of $\delta$ at the bounce point.  (On the other hand, if the total stress-energy tensor $\check T_{\mu\nu}$ has positive energy density, then $\delta \le 1$.)

It is now possible to derive transition rules for the quasi-Kasner exponents, generalizing the results of Section~\ref{ZeldovichUniverse}.  From \eqref{KasnerEvolve}, it follows that
\be
\frac{\sgn(\bar H) \cdot (k_i - \frac{1}{3})}{\delta} =  \frac{\lambda_i }{\sqrt{3\Sigma^2}}= {\rm constant}~,
\ee
and therefore, denoting by $\delta_-$ and $\delta_+$ respectively the values of the radial parameter corresponding to the contracting and the expanding branch,
\be
\delta_{+}\left(k_i^- -\frac{1}{3}\right)+\delta_{-}\left(k_i^+ -\frac{1}{3}\right)=0 ~.
\ee
Note that in this equation both $\delta$ and the quasi-Kasner exponents $k_i$ depend on time, even away from the bounce.  This can be rearranged to give
\be\label{TransitionRuleGeneral}
k_i^{+}=\frac{1}{3}\left(1+\frac{\delta_{+}}{\delta_{-}}\right)-\left(\frac{\delta_{+}}{\delta_{-}}\right)k_i^{-} ~,
\ee
which generalizes Eq.~\eqref{TransitionRule} established earlier for the transition between two Zeld'ovich universes.

The specific relation between $\delta_-$ and $\delta_+$ will depend on the specific modified gravity theory and matter content.  We will now consider two simple examples where the transition can be studied analytically, while a more general example is considered in detail in Appendix~\ref{SharpTransitions}.

As a first explicit example, it is possible to compute the evolution of the time-dependent Kasner exponents in the limiting curvature mimetic gravity theory~\cite{Chamseddine:2016uef}, which satisfies Conditions (i) and (ii).  (Note that while limiting curvature mimetic gravity exactly reproduces the LQC effective dynamics for a spatially flat FLRW model \cite{Bodendorfer:2017bjt, Liu:2017puc}, the dynamics are different for the Bianchi~I space-time \cite{Bodendorfer:2018ptp}.)  For limiting curvature mimetic gravity, the modified Friedmann equation obeyed by the mean scale factor is 
\be \label{mimetic-eq}
\bar{H}^2=\frac{1}{3} \left(\rho + \frac{\Sigma^2}{V^2} \right) \left(1-\frac{\rho}{\rho_c} - \frac{\Sigma^2}{\rho_c V^2} \right) ~,
\ee
and, considering the vacuum case for simplicity, the solution is \cite{Chamseddine:2016uef}
\be
\bar{a}(t)=\left(\frac{\Sigma^2}{\rho_c}\right)^{1/6}(1+3t^2\rho_c)^{1/6} ~.
\ee
From Eq.~\eqref{KasnerEvolve}, the time-dependent quasi-Kasner exponents are
\be \label{KasnerEvolution_Mimetic}
k_i(t)=\frac{1}{3}-\sgn(\bar{H})\left(k_i^{-} -\frac{1}{3}\right)\sqrt{1+\frac{1}{3\rho_c t^2}} ~.
\ee
Note that at $t=0$ the quasi-Kasner exponent diverges, but is finite at all other times.  It is easy to check that by defining $\lim_{t\to\pm\infty} k_i(t) = k_i^{\pm}$, the asymptotic vacuum transition rule $k_i^+ = \frac{2}{3} - k_i^-$ first derived in \cite{Chamseddine:2016uef} is recovered.  In the presence of matter with a generic equation of state, solving the effective Friedmann equation \eqref{mimetic-eq} is a more challenging task. Nonetheless, it remains possible to integrate the equations around the bounce~\cite{deCesare:2019pqj}; and then the time-dependent quasi-Kasner exponents are readily computed.

A more general example is given by the Kasner transition for a simple bounce model where the mean scale factor evolves as
\be \label{example-a}
\bar a(t) = \frac{a_o}{2} \, \varUpsilon t^2 + a_o~,
\ee
during the time interval $t \in [-T/2 + \tau, T/2 + \tau]$, with the bounce occurring at $t=0$.  Here $\varUpsilon$ (the value of $\dot{\bar H}$ at the bounce), $T$ (the duration of the bounce), and $a_o$ are all positive constants, while the parameter $\tau$ can be either positive or negative and measures the time asymmetry of the bounce.  In the following, we assume $|\tau|/T \ll 1$ and $\varUpsilon T^2 \ll 1$.  A more general class of bouncing models is examined in Appendix~\ref{AppendixBounce}.

The angular parameters $Z_i$ corresponding to the transition are determined by Eq.~\eqref{AngularVarTransition}, while the ratio $\delta_{+}/\delta_{-}$ is fixed by the time asymmetry $\tau/T$ (for intermediate steps in the calculation, see Appendix~\ref{AppendixBounce})
\be \label{DeltaRatioBounce}
\frac{\delta_{+}}{\delta_{-}} \simeq 1 - \frac{4\tau}{T}~,
\ee
where $\delta_{-}$  and $\delta_{+}$ denote the values of the parameter $\delta$ at the onset and at the exit of the bounce phase, repectively. Note that for a symmetric bounce where $\tau=0$, $\delta_{+}/\delta_{-} \simeq 1$ regardless of the matter content; this result holds for any bounce whose dynamics are described by \eqref{example-a} with $\varUpsilon T^2 \ll 1$.  This shows that the simplest type of generalized Kasner transitions studied in Section~\ref{ZeldovichUniverse} can also be applied more generally in the presence of other types of matter fields.  Away from the bounce, $\delta$ evolves according to Eq.~\eqref{ODE_CircleRadKasnerPlane2}.

\subsection{Geometric depiction of the quasi-Kasner exponent dynamics}

In addition, it is possible to geometrically describe the dynamics of the quasi-Kasner exponents.

This can be done by compactifying the Kasner plane (where $\sum_{i=1}^3 k_i = 1$) to a sphere by means of the mapping $\delta=\tan\frac{\theta}{2}$, with $0\leq\theta\leq\pi$. The isotropic point $k_1 = k_2 = k_3 = \frac{1}{3}$ lies on the north pole of the sphere, while the south pole corresponds to $\delta\to+\infty$. The three lines $k_i=0$ are mapped to the curves on the sphere with equation $Z_i(\psi)+\frac{1}{2}\cot(\frac{\theta}{2})=0$. Note that, when both matter and the effective fluid $\tilde T_{\mu\nu}$ have positive energy density, only the northern hemisphere can be attained by solutions satisfying the Einstein equations. The southern hemisphere is only accessible to modified gravity theories; in particular, the bounce point $\bar H = 0$ occurs at the south pole.

The evolution of the quasi-Kasner exponents is described by the motion of a point particle on the sphere; due to Eq.~\eqref{AngularVarTransition} this motion is constrained to take place along great circles on the sphere, passing through the poles.  As a result, on the sphere the generalized Kasner transition caused by the bounce corresponds to mapping meridians to their antimeridians.  (Projecting these meridians onto the Kasner plane gives the half-lines $r^-$ and $r^+$ corresponding to constant $\psi^-$ and $\psi^+$ discussed earlier.)

\begin{figure}[t]
\begin{center}
\includegraphics[width=0.5\columnwidth]{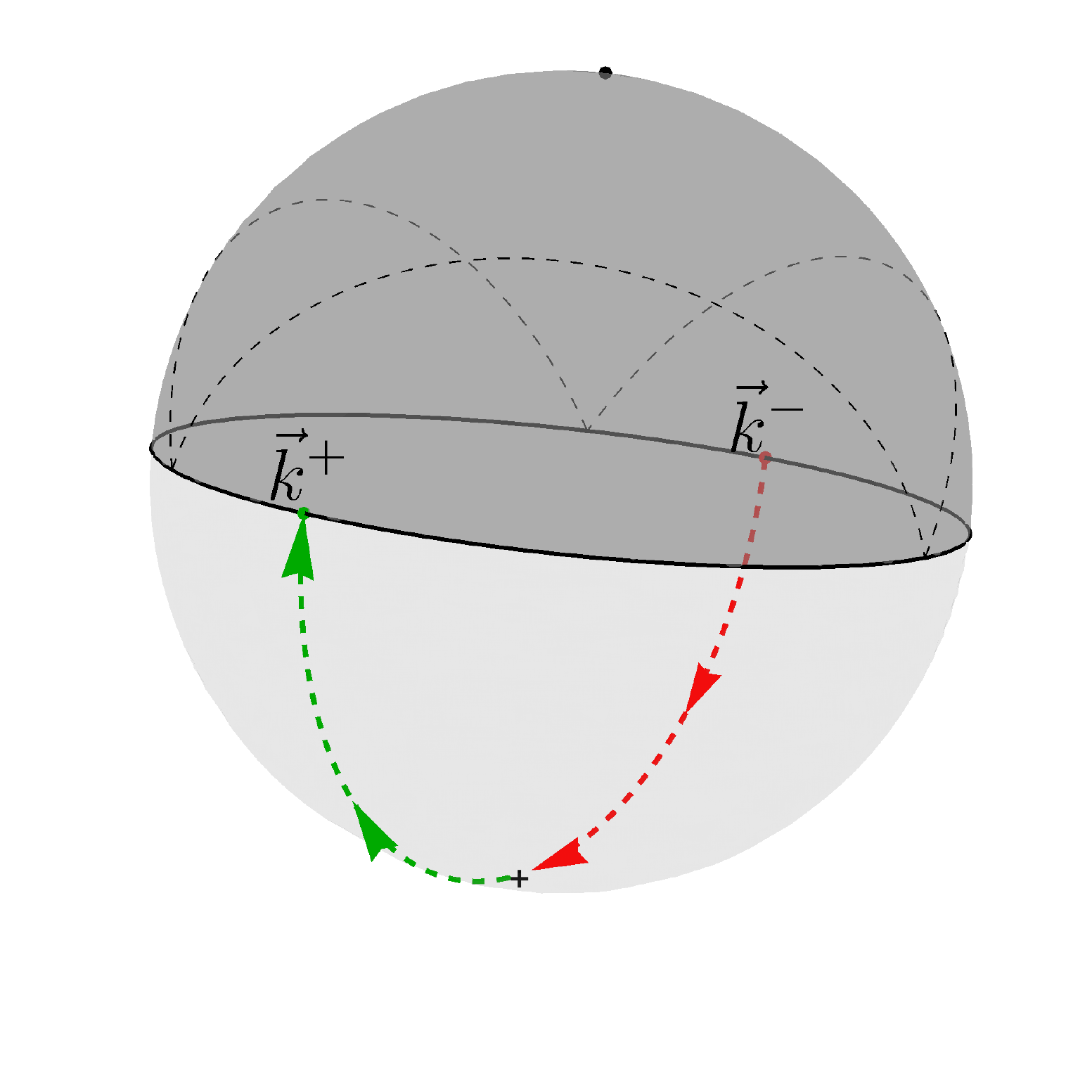}
\caption{\hangindent=12pt \small
The figure shows the one-point compactification of the Kasner plane. The Kasner circle is mapped to the equator and its interior is mapped the northern hemisphere, with the north pole representing a flat FLRW universe. The interior of the deformed triangle represents the region where all Kasner exponents are positive, cf.~Fig.~\ref{KasnerDisc}. Note that only the northern hemisphere is accessible if $\check T_{\mu\nu}$ has a positive energy density. However, the southern hemisphere becomes accessible during the bounce phase, where the energy density of the effective fluid becomes negative; the bounce itself occurs when the quasi-Kasner exponents are located at the south pole. The arrows illustrate the generalized Kasner transition for a bouncing vacuum Bianchi~I universe. When matter fields are present, the entire trajectory of the quasi-Kasner exponents still lies on the same great circle intersecting the north and south poles, but the end points are not constrained to lie on the equator, see Section~\ref{Sec3}. In particular, as shown in Section~\ref{GeneralTransition}, in the presence of matter with equation of state $\omega\neq1$ the end-points undergo further evolution both in the pre-bounce and in the post-bounce phases.}\label{sphere}
\end{center}
\end{figure}

An example of this trajectory is shown in Fig.~\ref{sphere} for the vacuum case where the quasi-Kasner exponents begin and end on the Kasner circle.  In the presence of matter fields, the start and end points will lie in the upper hemisphere, potentially at different heights, but necessarily on the same great circle passing through the north and south poles.  The relation between the co-latitudes $\theta$ of the initial and final states can be determined by solving Eq.~\eqref{ODE_CircleRadKasnerPlane2}.

Also, from this example it is clear that the introduction of quasi-Kasner exponents is also useful to describe the dynamics during the bounce for the vacuum and stiff matter cases: even though for these space-times in general relativity the Kasner exponents are constant, this is no longer the case in modified theories of gravity.

\section{Conclusions}

For a large class of modified gravity theories that produce a non-singular bounce---specifically, the theories that satisfy Conditions (i) and (ii) listed in the Introduction---it is possible to relate the pre-bounce and post-bounce Bianchi~I space-times through a set of linear transition rules relating their (quasi-)Kasner exponents, based on the conservation of the densitized shear.  In the vacuum case or in the presence of stiff matter (assuming the momentum of the scalar field is conserved by the bounce, otherwise see Eq.~\eqref{TransitionRule}), the asymptotic transition rules---i.e., far from the bounce where the modifications to general relativity are negligible---for the Kasner exponents are simply $k_i^+ = \frac{2}{3} - k_i^-$.  This result is the same as what has been found previously for limiting curvature mimetic gravity \cite{Chamseddine:2016uef}, which satisfied Conditions~(i) and~(ii), and for loop quantum cosmology \cite{Wilson-Ewing:2017vju}, which does not satisfy Condition~(i).

The fact that the generalized Kasner transition rule holds even for some theories that do not satisfy Condition (i) indicates that these transition rules likely hold much more generally.  In fact, if a non-singular bounce does not introduce a preferred direction, then there are only two possible transition rules that don't require a preferred direction: the identity map and the antipodal map \cite{Uggla} (see also a discussion on this point in Ref.~\cite{Wilson-Ewing:2018lyx} in terms of the Misner shape variables).  Here it is the antipodal map that is selected by the dynamics.

Also, by introducing the time-dependent quasi-Kasner exponents, it is possible to extend these results to include other types of matter content beyond the vacuum and stiff matter cases.  In addition, this provides a way to track the dynamics of the Bianchi~I space-time through the bounce, for all types of matter content.  The quasi-Kasner exponents always remain on the Kasner plane, and move radially towards or from the isotropic central point.  If the plane is compactified onto a sphere by including the point at infinity as the south pole, the dynamics of the quasi-Kasner exponents follow a meridian, passing through the south pole at the bounce point.

The evolution of the quasi-Kasner exponents is studied in detail for Bianchi~I space-times with multiple matter fields in Appendix~\ref{App.Seq} and through the bounce in Appendix~\ref{AppendixBounce}.  A brief study of models with multiple bounces as well as the emergent universe scenario is given in Appendix~\ref{MultiBounce}.

Finally, the line element for a Bianchi~I space-time with a stiff fluid and an ekpyrotic field with a constant equation of state $\omega = 3$ is presented in Sec.~\ref{ekpyrotic-sol}; to the best of our knowledge this is a new explicit solution to the Einstein equations.

\acknowledgments
We thank Alan Coley and Jean-Luc Lehners for helpful comments on an earlier draft of the paper.
This work was supported in part by the Natural Science and Engineering Research Council of Canada.
MdC also acknowledges support from the Atlantic Association for Research in the Mathematical Sciences.

\appendix

\section{Evolution of anisotropies in general bouncing models}
\label{SharpTransitions}

\subsection{Multiple matter fields}
\label{App.Seq}

In many cosmological models there are several different matter fields present, which typically dominate the dynamics at different times.  In the following, we assume that the dynamics of the space-time can be split into different eras (with instantaneous transitions between eras), each of which can be approximated as being dominated by one matter field only.  Then, in each era, the directional Hubble rates are determined by the energy density and pressure of the dominant matter field only, with the directional scale factors and Hubble rates being continuous during the transition between eras dominated by different matter fields.  (Note that corrections to general relativity captured in $\tilde T_{\mu\nu}$ are treated as an effective fluid, which will typically dominate the dynamics during the bounce.)

For definiteness, the time of the $r^{\rm th}$ transition between different matter fields dominating the dynamics will be denoted as $t_r$. The initial time $t_0$ is set in the contracting phase. The bounce takes place at $t_p$. The instants of time when the universe enters and exits the bounce phase are denoted by $t_{p-1}$ and $t_{p+1}$, respectively. In the time interval $[t_{r-1},t_{r}]$, the space-time is assumed to be dominated by a fluid with constant equation of state~$\omega_r$.  (The generalization to a varying equation of state is direct, although in general it will not be possible to find analytic solutions.) It is convenient to introduce the parameter $\epsilon_r=\frac{3}{2}(\omega_r+1)$. For $\epsilon_r\neq0$, the evolution of the mean scale factor in the time interval $t \in [t_{r-1}, t_r]$ is given by
\be\label{ScaleFactorApproxEvol1}
\bar{a}(t)=\bar{a}_r\left(\frac{t-\tilde{t}_r}{t_r-\tilde{t}_r}\right)^{\frac{1}{\epsilon_r}} ~,
\ee
while if $\epsilon_r=0$ (i.e., $\omega_r=-1$, corresponding to a universe dominated by a positive cosmological constant), then
\be\label{ScaleFactorApproxEvol2}
\bar{a}(t)=\bar{a}_r\exp\left(\frac{t-\tilde{t}_r}{t_r-\tilde{t}_r}\right) .
\ee
In these expressions, the integration constant $\tilde{t}_r$ determines the value of the mean Hubble rate at $t=t_r$, and is also related to the mean Hubble rate at $t=t_{r-1}$,
\be
\bar{H}(t_r) = 
\begin{cases}
\Big(\epsilon_r(t_r-\tilde{t}_r)\Big)^{-1} = \Big(\epsilon_{r+1}(t_r-\tilde{t}_{r+1})\Big)^{-1}~,~  & {\rm if}~\epsilon_r\neq0~, \\
(t_r-\tilde{t}_r)^{-1}~,~ & {\rm if}~\epsilon_r=0  ~.  \\
\end{cases}
\ee
By evaluating $\bar a(t_r)$ using Eq.~\eqref{ScaleFactorApproxEvol1} for the time intervals $[t_{r-1}, t_r]$ and $[t_r, t_{r+1}]$, it follows that (if $\epsilon_r \neq 0$)
\be
\frac{\bar{a}_r}{\bar{a}_{r-1}}=\left(\frac{\bar{H}(t_{r-1})}{\bar{H}(t_{r})}\right)^{1/\epsilon_r}~,
\ee
which in turn implies
\be
\frac{\Omega_{\Sigma}(t_r)}{\Omega_{\Sigma}(t_{r-1})}=\left(\frac{\bar{a}_{r}}{\bar{a}_{r-1}}\right)^{2(\epsilon_r-3)} ~.
\ee
Note that this last relation also holds for $\epsilon_r = 0$, even though the intermediate step does not.

Introducing the number of e-folds of expansion in the $r^{\rm th}$ era
\be
N_r \coloneqq \log\left(\frac{\bar{a}_r}{\bar{a}_{r-1}}\right) ~,
\ee
(by definition, $N_r>0$ in the expanding phase, while $N_r<0$ in the contracting phase), and this definition combined with the previous result gives
\be
N_r = \frac{1}{2(\epsilon_r-3)}\log \left(\frac{\Omega_{\Sigma}(t_r)}{\Omega_{\Sigma}(t_{r-1})}\right) ~.
\ee
This relation holds for $\epsilon_r \neq 3$, while if $\epsilon_r = 3$ then $\Omega_{\Sigma}(t_r) = \Omega_{\Sigma}(t_{r-1})$).

To calculate the evolution of the anisotropies, governed by Eq.~\eqref{SolAlpha}, it is convenient to subtract the isotropic part of the logarithmic scale factors.  Introducing $\beta_i=\alpha_i-\frac{1}{3}\sum_j\alpha_j$, Eq.~\eqref{SolAlpha} becomes
\be
\beta_i=\beta_i^0 + \lambda_i \int_{t_0}^t \frac{\de t'}{V(t')}~.
\ee
(Note that $\beta_i^0=\alpha_i^0$ since, by definition, $\sum_{i=1}^3\alpha_i^0=0$.)  The integral can be split into two parts,
\be
\beta_i(t) = \beta_i^0+\lambda_i\left(\sum_{s=1}^{r} I_{s}+ \int_{t_r}^t \frac{\de t'}{V(t')}\right) ~,
\ee
where $t_r$ is the last transition between different matter fields dominating the dynamics before the time $t$.  Here $I_j$ is defined as
\be
I_s = \int_{t_{s-1}}^{t_s}\frac{\de t'}{V(t')}~.
\ee
Note that $I_s > 0$, since $V(t) > 0$ and by definition $t_s > t_{s-1}$.

In the pre-bounce phase ($s < p-1$) and in the post-bounce phase ($s > p+1$), the integrals $I_s$ are easily evaluated from~\eqref{ScaleFactorApproxEvol1} or \eqref{ScaleFactorApproxEvol2}.  For $\epsilon_s \neq 3$ (and including the case $\epsilon_s = 0$),
\begin{align}
I_s = \frac{1}{\epsilon_s-3} \cdot \frac{1}{\bar{H}(t) V(t)} \Big|_{t_{s-1}}^{t_s}
&= \frac{ \sgn\big(\bar{H}(t_s)\big)} {\epsilon_s-3} \sqrt{\frac{3}{\Sigma^2}} \cdot \sqrt{\Omega_\Sigma} \Big|_{t_{s-1}}^{t_s} \nonumber \\
&=\frac{\sgn\big(\bar{H}(t_s)\big)}{\epsilon_s-3}\sqrt{\frac{3 \Omega_{\Sigma}(t_{s-1})}{\Sigma^2}} \Big(\exp\big(N_s(\epsilon_s-3)\big)-1\Big) ~, \label{ResultIntegralGeneral}
\end{align}
while for $\epsilon_s = 3$, in which case $\Omega_{\Sigma}(t_s) = \Omega_{\Sigma}(t_{s-1})$,
\be
I_s = \frac{N_s}{\bar{H}(t_s) V(t_s)} = \sgn\big(\bar{H}(t_s)\big) \sqrt{\frac{3 \Omega_{\Sigma}(t_s)}{\Sigma^2}} N_s ~.\label{ResultIntegralStiff}
\ee
Note that this result equals $\lim_{\epsilon_s \to 3} I_s$ in~\eqref{ResultIntegralGeneral}, and therefore $I_s$ is a continuous function in terms of $\epsilon_s$.

From these results, the value of $\Omega_{\Sigma}(t_r)$, with $t_r$ the time at the end of the $r^{\rm th}$ era before the bounce (i.e., $r\leq p-1$), is
\be\label{EvolveDensityBeforeBounce}
\Omega_{\Sigma,r}= \frac{\Omega_{\Sigma,r}}{\Omega_{\Sigma,r-1}} \cdot \frac{\Omega_{\Sigma,r-1}}{\Omega_{\Sigma,r-2}} \cdot \ldots \cdot \frac{\Omega_{\Sigma,1}}{\Omega_{\Sigma,0}} \cdot \Omega_{\Sigma,0} = \Omega_{\Sigma,0} \cdot \exp\left(2\sum_{s=1}^r N_s (\epsilon_s-3) \right) ~,
\ee
where $\Omega_{\Sigma,s} = \Omega_\Sigma(t_s)$, while for the $r^{\rm th}$ era after the bounce,
\be\label{EvolveDensityAfterBounce}
\Omega_{\Sigma,r} = \frac{\Omega_{\Sigma,r}}{\Omega_{\Sigma,r-1}} \cdot \frac{\Omega_{\Sigma,r-1}}{\Omega_{\Sigma,r-2}} \cdot \ldots \cdot \frac{\Omega_{\Sigma,p+2}}{\Omega_{\Sigma,p+1}} \cdot \Omega_{\Sigma,p+1} =\Omega_{\Sigma,p+1} \cdot \exp\left(2\sum_{s=p+1}^r N_s (\epsilon_s-3) \right) ~,
\ee
with in this case $r \geq p+1$.  In terms of $\delta_s = \sqrt{\Omega_{\Sigma,s}}~$,
\be
\delta_r = 
\begin{cases}
\delta_0 \cdot \exp\Big(\sum_{s=1}^r N_s (\epsilon_s-3) \Big) ~,~~ & \mbox{if $r\leq p-1$} ~, \\
\delta_{p+1} \cdot \exp\Big(\sum_{s=p+1}^r N_s (\epsilon_s-3)\Big) ~,~~& \mbox{if $r\geq p+1$} ~. \\
\end{cases}
\ee
The quasi-Kasner exponents stay on the same constant $\psi$ half-line in the Kasner plane, with the distance from the isotropic point $k_1 = k_2 = k_3$ given by $\sqrt{\frac{2}{3} \,} \delta$.

The remaining task is to evaluate how $\Omega_\Sigma$ (and $\delta$ and $k_i$) evolve across the bounce.

\subsection{The bounce}
\label{AppendixBounce}

The evolution of the anisotropies across the bounce is encoded in
\be
I_B = \int_{t_{p-1}}^{t_{p+1}} \frac{\de t'}{V(t')}~.
\ee
To evaluate this integral, it is necessary to know the dynamics of $\bar a(t)$.  First, without any loss of generality we choose the bounce time to be $t_p=0$.

We will assume that during the bounce the mean Hubble rate behaves as
\be \label{bounce-evol}
\bar{H}(t)\simeq\varUpsilon\, t^{2n-1}+\Phi\, t^{2n} ~,
\ee
where $n$ is a positive integer and is called the order of the bounce \cite{Cattoen:2005dx}; this is a generalization of the (first-order) bounce models often considered \cite{Cai:2012va, Lin:2010pf}.  Note that the condition $\dot{\bar H}(t=0) > 0$ implies that $\varUpsilon>0$.

In addition, denoting the duration of the bounce by $T=t_{p+1}-t_{p-1}$ and its asymmetry by $\tau=\frac{1}{2}(t_{p+1}+t_{p-1})$, we will assume that $\tau\ll T$, and that the bounce is `fast', i.e., that
\be\label{FastBounceConstraints}
\varUpsilon\, T^{2n}\ll1~,\hspace{1em} |\Phi|\, T^{2n+1}\ll1 ~.
\ee
Note that besides the duration, a cosmological bounce is also characterized by the timescale $\bar H_{\rm m}^{-1}$, the inverse of the maximal value of the Hubble rate $\bar{H}_{\rm m} = \max \{ H\left(\frac{T}{2}\right),-H\left(-\frac{T}{2}\right)\}$; the fast bounce condition implies that $T\ll \bar{H}_{\rm m}^{-1}$.  Note that the fast bounce condition is independent of the order $n$ of the bounce.

The final assumption here is that the Hubble rate is strictly increasing during the bounce phase (this assumption is relaxed in Appendix~\ref{MultiBounce}), which implies $|\Phi| T/ 2\varUpsilon \le (2n-1)/2n$.  If $\Phi \neq 0$, this can be rewritten as
\be \label{maxT}
T \le \frac{2\varUpsilon}{|\Phi|} \left(1-\frac{1}{2n}\right)~.
\ee

With these assumptions, it is possible to calculate how the anistropies evolve across the bounce.  First, the expression for $\bar H(t)$ can be integrated, giving
\be
\bar{a}(t)\simeq\bar{a}_{B}\exp\left(\frac{\varUpsilon}{2n}\, t^{2n}+\frac{\Phi}{2n+1}\, t^{2n+1} \right)\simeq \bar{a}_{B} \left( 1+\frac{\varUpsilon}{2n}\, t^{2n}+\frac{\Phi}{2n+1}\, t^{2n+1}\right) ~,
\ee
where the last approximation is justified by the fast bounce conditions.

The number of e-folds of expansion during the bounce is
\be
N_B = \log \left( \frac{a_{p+1}}{a_{p-1}} \right) = \frac{\varUpsilon}{2n}\, \left( t^{2n}_{p+1} - t^{2n}_{p-1} \right) +\frac{\Phi}{2n+1}\, \left( t^{2n+1}_{p+1} - t^{2n+1}_{p-1} \right) ~,
\ee
a symmetric bounce will have $N_B = 0$. The fast bounce condition implies $| N_B | \ll 1$, and to leading order,
\be\label{EfoldsBounce}
N_B \simeq \frac{2A\varUpsilon}{2n+1}\left(\frac{T}{2}\right)^{2n} ~,
\ee
where $A = \Phi T/ 2\varUpsilon$, and $|A| < 1-\frac{1}{2n}$ by~\eqref{maxT} (although note that this does not imply $|A| \ll 1$; the smallness of $N_B$ follows solely from the fast bounce assumption).

Evaluating the integral $I_B$ to leading order gives
\be\label{IntegralFastBounceExpand}
I_B = \int_{t_{p-1}}^{t_{p+1}}\de t\; \bar{a}(t)^{-3}\simeq\bar{a}_{B}^{-3}\, T\left(1-\frac{3\,\varUpsilon}{2n(2n+1)}\left(\frac{T}{2}\right)^{2n}\right)\simeq\bar{a}_{B}^{-3}\, T\left(1-\frac{3}{4n A}N_B\right) ~.
\ee
Note that both in \eqref{EfoldsBounce} and in \eqref{IntegralFastBounceExpand}, the $\Phi$ term is subleading since it appears in combination with $\frac{\tau}{T} \ll 1$; this is true even if $\Upsilon T^{2n}$ and $|\Phi|\, T^{2n+1}$ are similar in magnitude.
In general, $\bar H(t_{p-1}) \neq - \bar H(t_{p+1})$. To first order in $\tau/T$, their ratio is given by
\be
\frac{\bar H(t_{p+1})}{\bar H(t_{p-1})}=-\left(\frac{1+A}{1-A}\right)\left[1+4\left(2n-\frac{1}{1-A^2}\right)\cdot\frac{\tau}{T} \right]+\mathcal{O}\left(\frac{\tau}{T}\right)^2 ~,
\ee
and as a result the density parameters before and after the bounce are related by
\be\label{OmegaEvolveGeneralBounce}
\frac{\Omega_{\Sigma}(t_{p+1})}{\Omega_{\Sigma}(t_{p-1})} = \left(\frac{H(t_{p-1})}{H(t_{p+1})}\right)^2 e^{-6N_B} \simeq \left(\frac{1-A}{1+A}\right)^2 \left[1-8\left(2n-\frac{1}{1-A^2}\right)\cdot \frac{\tau}{T} -6N_{B}\right] ~,
\ee
neglecting higher order terms. Interestingly, the leading order term does not depend on the order $n$ of the bounce.

Finally, the evolution of the Jacobs $\delta$ parameter through the bounce follows from $\delta_{r} = \sqrt{\Omega_{\Sigma}(t_r)}$~.  The maximum possible damping of anisotropies is obtained for $A = 1-\frac{1}{2n}$, in which case one finds to leading order
\be
\left(\frac{\delta_{p+1}}{\delta_{p-1}}\right)_{\rm min} \simeq \frac{1}{4n-1} ~,
\ee
whereas the maximum amplification is obtained for $A = -1 + \frac{1}{2n}$, to leading order
\be
\left(\frac{\delta_{p+1}}{\delta_{p-1}}\right)_{\rm max} \simeq 4n-1 ~.
\ee
Thus, even for a low order bounce (i.e., $n\sim1$), a significant damping or amplification of the anisotropies can occur during the bounce if the parameters $\varUpsilon$, $\Phi$, $T$ are such that $|A| \sim 1 - \frac{1}{2n}$.

During the bounce, $\delta$ diverges at $t=0$ (since $\bar H(t=0) = 0$), at this time the quasi-Kasner exponents are not defined.  However, at all other times the quasi-Kasner exponents are well defined and move on a half-line in the Kasner plane defined by the constant Jacobs angle $\psi$, with a distance ${\sqrt\frac{2}{3}\,}\delta(t)$ from the isotropic point $k_1 = k_2 = k_3 = \frac{1}{3}$.  Then, after the bounce $\sgn(\bar H)$ changes sign, implying $\psi^+ = \psi^- + \pi$.

If $\bar H(t)$ is monotonic during the bounce (as assumed in the calculations here), then $\delta(t)$ increases monotonically until the bounce point, at which point $\delta(t)$ diverges; once the space-time starts expanding, then $\delta$ decreases from infinity.  As for the quasi-Kasner exponents, as depicted in Fig.~\ref{sphere}, they start on a half-line of constant $\psi^-$, and then at the bounce are mapped to the other half-line of constant $\psi^+ = \psi^- + \pi$ and travel inwards towards the isotropic point (assuming the dominant matter field is not ekpyrotic).  The precise end point for the quasi-Kasner exponents is determined by $\lim_{t \to +\infty} \delta(t)$, which depends on the matter content of the space-time.

On the other hand, if the `bounce' era is not monotonic in $\bar H(t)$, then there must be multiple bounces and recollapses of the mean scale factor, this possibility is considered next.

\subsection{Generalizations beyond single-bounce cosmology}
\label{MultiBounce}

In the case that $\bar H$ is not monotonic during the time interval where $\tilde T_{\mu\nu}$ dominates the dynamics, there are necessarily multiple bounces and recollapses of the mean scale factor.  Since the evolution equation \eqref{EqForK_general} for the time-dependent quasi-Kasner exponents is completely general within the class of theories satisfying Condition (i), it also holds in this case.  As a result, it follows that $\delta(t)$ diverges at each bounce and also at each recollapse point, while the quasi-Kasner exponents in all contracting periods (when $\bar H < 0$) lie on the same $\psi^-$ half-line and lie on the same $\psi^+ = \psi^- + \pi$ half-line in all expanding periods (when $\bar H > 0$).  In addition, the transition rule \eqref{TransitionRuleGeneral} still applies to each bounce individually; moreover, a similar transition rule also holds for each recollapse that occurs between consecutive bounces.

To see this, note that the relations \eqref{KasnerEvolve} and \eqref{AngularVarTransition} hold in full generality, including around the recollapse point.  Therefore, solving for $\delta_\pm$ either side of the recollapse determines the quasi-Kasner exponents $k_i^\pm$.  As a result, the evolution of anisotropies in a Bianchi~I universe with multiple bounces and recollapses can be described as a sequence of generalized Kasner transitions \eqref{TransitionRuleGeneral}, which are completely specified once $\delta_{\pm}$ are determined for each bounce and for each recollapse.

\bigskip

Another interesting case is the emergent universe scenario, which has been studied in the context of string gas cosmology~\cite{Brandenberger:2008nx} and inflation in a closed FLRW space-time \cite{Ellis:2003qz}, and is similar to the Galilean genesis scenario~\cite{Creminelli:2010ba}. In this case the universe goes through a transition from a quasi-static phase in the past to an expanding phase in the future. The mean scale factor (and therefore the volume) is always monotonically increasing with $\bar{H}>0$ at all times, and in the past approaches a constant value as $t\to-\infty$, implying that $\lim_{t\to-\infty}\bar{H}=0$. Note that in such a scenario, $\lim_{t \to -\infty}\epsilon$ diverges and the quasi-Kasner exponents are not well-defined in this limit (this is exactly analogous to how the quasi-Kasner exponents diverge at the bounce point in non-singular bounce cosmologies).

In a particularly simple example of an emergent universe, the mean scale factor evolves as $\bar{a}(t)=a_0+A\,\mbox{e}^{H_0 t}$ \cite{Ellis:2003qz}, and then $\epsilon(t)=-a_0 A^{-1}\mbox{e}^{-H_0 t} < 0$. Assuming that Condition (i) holds, the dynamics of the quasi-Kasner exponents can be determined, given some initial conditions $\delta(t_o), \psi(t_o)$.  Since $\bar H > 0$, then $\sgn(\bar H)$ does not change and therefore $\psi(t)=\psi(t_o)$ is a constant.  The dynamics of $\delta$ are obtained by solving~\eqref{ODE_CircleRadKasnerPlane2}, which gives
\be
\delta(t) = c \: \frac{\mbox{e}^{-H_0 t}}{\bar{a}^2(t)}~.
\ee
Here $c$ is a constant fixed by the initial conditions. For $c=0$ the evolution is trivial, since $\delta=0$ and the universe is exactly isotropic at all times. The case $c\neq0$ is more interesting: $\delta$ diverges as $t$ tends to $-\infty$ and monotonically decreases as the universe starts expanding; in the asymptotic future, $\delta\approx  c\,\mbox{e}^{-3H_0 t}$ as in inflation and the universe is rapidly isotropized.  In this case, the evolution of the quasi-Kasner exponents can be represented geometrically as the motion of a point starting from the south pole of the sphere corresponding to the compactification of the Kasner plane, and moving to the northern hemisphere---this can be understood as the expanding phase, with an infinite duration, of a bounce.  Note that this qualitative behaviour for the dynamics of $\delta(t)$ in an emergent universe can be expected if the dynamics are generated by any modified gravity theory that satisfies Condition (i).

\section{The conservation law for $f(R)$ gravity in the Jordan frame}
\label{FofRgravity}

In (metric) $f(R)$ gravity the field equations for the theory formulated in the Jordan frame can be recast as
\be\label{Eq:EffectiveEinstein_FofR}
G_{\mu\nu}=\frac{1}{f'(R)}\left(T_{\mu\nu}+\tilde{T}_{\mu\nu}\right) ~,
\ee
where a prime denotes differentiation with respect to $R$, and the effective stress-energy tensor is \cite{Sotiriou:2008rp}
\be
\tilde{T}_{\mu\nu}=\frac{1}{2}\Big( f(R)-R \, f'(R) \Big) g_{\mu\nu}+\nabla_\mu\nabla_\nu f'(R)-  \Box f'(R)\,g_{\mu\nu}  ~.
\ee
In a Bianchi~I space-time,
\be
\nabla_\mu\nabla_\nu \Big( f'(R) \Big) = \Big(\pa_{t}^2 f'(R) \Big)\,u_\mu u_\nu - \Big(\pa_{t}f'(R) \Big) \, K_{\mu\nu} ~,
\ee
with $u_\mu$ the unit time-like normal to the spatial surfaces, and therefore
\begin{align}
\tilde{T}^0_{\pha 0}&=\frac{1}{2}\Big(f(R)-R f'(R)\Big)-K\Big(\pa_{t}f'(R) \Big) ~,\\
\tilde{T}^i_{\pha j}&=\left[\frac{1}{2}\Big(f(R)-R f'(R)\Big)-\Box f'(R) \right]\delta^i_{\pha j} + K^i_{\pha j} \Big( \pa_{t} f'(R) \Big) ~.
\end{align}
From Eq.~\eqref{Eq:EffectiveEinstein_FofR}, following the same steps used to derive Eq.~\eqref{SpaceSpace_TraceFree} gives
\begin{align}
-\frac{1}{\sqrt{h}}\frac{\de}{\de t}\left[\sqrt{h}\left( K^i_{\pha j}-\frac{1}{3}K \delta^i_{\pha j}\right)\right] &=\frac{1}{f'(R)}\left(\tilde{T}^i_{\pha j}-\frac{1}{3}\tilde{T}^k_{\pha k}\delta^i_{\pha j}\right) \nonumber \\ 
&=\frac{\Big(\pa_{t}f'(R)\Big)}{f'(R)}\left( K^i_{\pha j}-\frac{1}{3}K \delta^i_{\pha j}\right) ~,
\end{align}
and therefore, in $f(R)$ gravity in the Jordan frame there is the following conservation law in Bianchi~I space-times:
\be\label{ConservationLaw_FofR}
\frac{\de}{\de t}\left( \sqrt{h} \, f'(R) \, \sigma^i_{\pha j}\right) = 0~.
\ee
This conservation law is very similar to the conservation law~\eqref{ConservationLaw} for $f(R)$ gravity in the Einstein frame.

An entirely analogous conservation law holds in the general Brans-Dicke theory with a potential, provided that $f'(R)$ is replaced by the Brans-Dicke field $\phi$ in Eq.~\eqref{ConservationLaw_FofR}. In fact, since $f(R)$ gravity is a particular case of the general Brans-Dicke theory, the conservation rule for $f(R)$ gravity in the Jordan frame can be derived from the more general conservation law for Brans-Dicke theories in the Jordan frame.  Further, from this it is also clear that an analogous result also holds in Jordan-frame Palatini $f(R)$ gravity, since it is also a particular case of the family of Brans-Dicke theories \cite{Sotiriou:2008rp}.

From the conservation law \eqref{ConservationLaw_FofR}, the anisotropies evolve through the bounce as
\be\label{TransitionInFofR}
V_{+}\phi_{+}\dot{\beta}_i^{+}=V_{-}\phi_{-}\dot{\beta}_i^{-} ~,
\ee
where $\phi=f^{\prime}(R)$ in $f(R)$ gravity. This result implies that the transition rule for quasi-Kasner exponents has the same form as \eqref{TransitionRuleGeneral}, with
\be
\frac{\delta_{+}}{\delta_{-}}=\frac{\left| H_{-} \right| V_{-} \phi_{-}}{H_{+}V_{+}\phi_{+}}=
\frac{\phi_{-}}{\phi_{+}}\left(\frac{\Omega_{\Sigma}^{+}}{\Omega_{\Sigma}^{-}}\right)^{1/2} ~.
\ee
As usual, $\psi^+ = \psi^- + \pi$, but the evolution of the radial Jacobs parameter $\delta$ is different.  This is due to the presence of the Brans-Dicke field $\phi$, which determines the strength of the gravitational interaction in the Jordan frame. In the case of $f(R)$, $\phi$ depends on the shear through $R$, since
\be
R=-6\left(2\bar{H}^2+\dot{\bar{H}}\right)-\sum_i(\dot{\beta}_i)^2 ~.
\ee
Thus, in the Jordan frame there is a similar generalized Kasner transformation rule, although with a different relation between $\delta_+$ and $\delta_-$ due to the varying gravitational coupling $\phi^{-1}$.

\raggedright

\end{document}